%% file: main.tex
\newif\ifanonymized
\anonymizedfalse    %

\newif\ifsubmit
\submitfalse	%
\submittrue	%

\documentclass[11pt]{article}
   \usepackage[text={6.5in,9in},centering]{geometry}
   \usepackage{times}
\input{preamble}

\usepackage{setspace}
\usepackage[font={small,stretch=1},labelfont=bf]{caption}

\begin{document}

\title{Building an Open-Source Community to Enhance Autonomic Nervous System Signal Analysis: DBDP-Autonomic}
\date{}

\ifanonymized
    \author{Anonymized for blind submission}
\else
    \author{Jessilyn Dunn$^{1+}$, Varun Mishra$^{2+}$\footnote{contact author: v.mishra@northeastern.edu} , Md Mobashir Hasan Shandhi$^{1}$, Hayoung Jeong$^{1}$,\\ Natasha Yamane$^{2}$, Yuna Watanabe$^{2}$, Bill Chen$^{1}$, Matthew S. Goodwin$^{2}$\\
     $^{1}$ Duke University, $^{2}$ Northeastern University\\
$^{+}$co-first authors
}
\fi

\maketitle  %

\input{abstract}		%

\input{headings}	%

\input{body}		%

\input{ack}			%

\linespread{1}

\bibliographystyle{plainDOI}

\begingroup
\setstretch{1.0}
\bibliography{main}
\endgroup

\end{document}

%% file: preamble.tex
\usepackage{fancyhdr}

\usepackage{url}

\usepackage{color}

\usepackage[mmddyyyy,hhmmss]{datetime}

\usepackage{xspace}

\usepackage{graphicx}
\DeclareGraphicsExtensions{.pdf,.png,.eps,.ps,.jpg}

\ifsubmit
  \newcommand{\hey}[1]{\relax}
  \newcommand{\heyvarun}[1]{\relax}
  \newcommand{\bibtex}[1]{\relax}
  \newcommand{\note}[1]{\relax}
\else
  \newcommand{\hey}[1]{\textcolor{magenta}{[{#1}]}}
  \newcommand{\heyvarun}[1]{\textcolor{blue}{[Varun: {#1}]}}
  \newcommand{\note}[1]{\par\textcolor{magenta}{Note: {#1}}\par}
  \newcommand{\bibtex}[1]{\textcolor{red}{@bibtex}\{#1\}} 
\fi

\newcommand{\hide}[1]{\relax}

\newcommand{\figlabel}[1]{\label{fig:#1}}
\newcommand{\figref}[1]{Figure~\ref{fig:#1}}

\newcommand{\addfigure}[3]{
\begin{figure}[tb]
\centerline{\resizebox{#2\linewidth}{!}{\includegraphics{figs/#1}}}
\caption{#3}
\figlabel{#1}
\end{figure}
}

\newcommand{\seclabel}[1]{\label{sec:#1}}

%% file: abstract.tex
\begin{abstract}
Smartphones and wearable sensors offer an unprecedented ability to collect peripheral psychophysiological signals across diverse timescales, settings, populations, and modalities. However, open-source software development has yet to keep pace with rapid advancements in hardware technology and availability, creating an analytical barrier that limits the scientific usefulness of acquired data. We propose a community-driven, open-source peripheral psychophysiological signal pre-processing and analysis software framework that could advance biobehavioral health by enabling more robust, transparent, and reproducible inferences involving autonomic nervous system data. 
\end{abstract}

%% file: headings.tex
\ifsubmit
    \relax
\else
    \par\noindent \textcolor{red}{\textbf{DRAFT}: \today\ -- \currenttime}
    \pagestyle{fancy}
    \lhead{DRAFT in preparation}
    \rhead{version: \today\ -- \currenttime}
    \chead{}
    \lfoot{}
    \rfoot{}
\fi

%% file: body.tex
\section{Introduction} %
\seclabel{introduction}

Chronic physical conditions and mental health disorders are increasingly prevalent~\cite{worldhealthorganization:Worldhealthstatistics-}. Integrating biological, behavioral, and environmental data is needed to enable early detection, just-in-time intervention, and outcome monitoring to promote biobehavioral health.

Mobile monitoring technologies are changing how we collect behavioral, environmental, and peripheral psychophysiological data in non-clinical settings. They allow researchers and clinicians to gather information continuously and unobtrusively beyond traditional clinical settings' temporal and spatial limitations. While there have been successful examples of technology-driven biobehavioral applications such as detecting stress by continuously monitoring heart rate variability~\cite{hovsepian:cstress,mishra:continuous-stress}, predicting episodes of depression~\cite{chikersal:depression,wang:depression18}, or identifying sleep disorder patterns using wearable sleep trackers~\cite{dezambotti:WearableSleepTechnology-2019}, the full potential of these technologies has yet to be realized. There are at least two key challenges hindering progress: (1) the common misinterpretation of peripheral psychophysiological signals in biobehavioral research and (2) the need for greater transparency and reproducibility in biobehavioral research that involves peripheral physiological data.

\subsection{The Criticality of Context in the Interpretation of Autonomic Nervous System Data}
The Autonomic Nervous System (ANS) plays a fundamental role in regulating myriad physiological processes within the body, including, but not limited to, heart rate, blood pressure, respiration, and digestion. Its primary function is maintaining homeostasis by continuously adjusting the body to changing internal and external conditions. Thus, understanding the context in which ANS data is collected is critical to making accurate and meaningful biobehavioral interpretations. 

The ANS is highly variable within and across people but also has stable characteristics based on an individual’s baseline biology. ANS activity can be influenced by various factors, including but not limited to, stress, affect, cognition, physical activity, sleep, illness, medications, and environmental demands. ANS activity also varies over time—for example, heart rate variability changes during different stages of sleep~\cite{vanoli:HeartRateVariability-1995}. Further, ANS responses differ between individuals based on age, genetics, and health conditions~\cite{dart:Gendersexhormones-2002,moodithaya:GenderDifferencesAgeRelated-2011}. The combination of these influencing factors makes the ANS particularly difficult to study, especially when the broader context (setting, activity, health status, etc.) and a person’s baseline state are not considered.

Fluctuations in ANS signals may be misattributed to one factor (e.g., a stressful event) when they are the result of another factor (e.g., physical activity) (\figref{stress-example}). Some common examples of transient states that are misinterpreted when context is excluded include affect (emotion/mood~\cite{siegel:Emotionfingerprintsemotion-2018}), cognition (challenge/threat~\cite{wright:Cardiovascularcorrelateschallenge-2003}), and physical perturbations (sleep, medications, exercise~\cite{mishra:context}). This impacts broader digital biomarker development that focuses on a single physiological system and thus ignores the broader context and systemic interconnectedness that may collectively influence diagnostic outcomes (autonomic neuropathy, neurodegenerative disease, gastrointestinal disorders, etc.). Context also helps account for psychophysiological differences by differentiating between a stimulus-driven or condition-specific shift in ANS activity versus natural fluctuations (e.g., diurnal variations) around a baseline. Additionally, including context in research and clinical care enables a more quantitative assessment of the effects of interventions like medications and therapies and whether observed changes constitute lasting outcomes, which are notoriously difficult to assess systematically. 

\addfigure{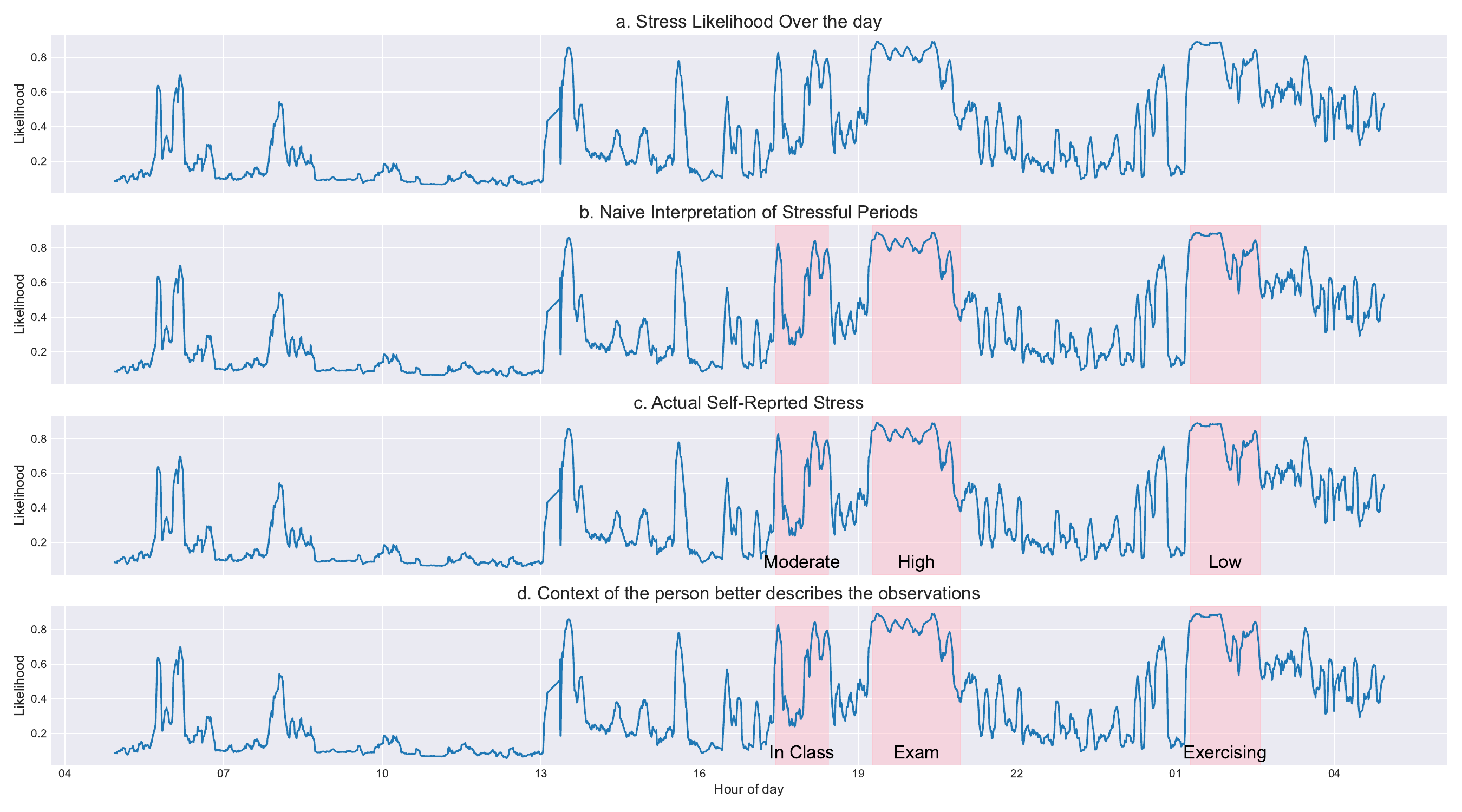}{1}{This figure demonstrates the need for additional context when analyzing ambulatory physiological signals. We used a stress-prediction model on heart rate variability data to predict a probability of physiological stress/arousal for a person over 24 hours. From a naïve interpretation, it seems there are three major stressful (or high arousal) periods (Fig. 1b). However, when asked to self-report stress levels, the user rated a mix of low to high stress for those periods, contradictory to the purely physiological interpretation (Fig. 1c). Considering additional contextual information, we realize that only one high-arousal episode was stressful since the user was undergoing an exam. The other periods were when the user was in a class and exercising later during the day, which showed similar physiological arousal but were not stressful (Fig. 1d).}

In sum, it is important to consider internal, external, and temporal contexts to ensure accurate diagnosis and interpretation of ANS using digital health data. Software that enables data fusion and multi-modal analysis is needed to address these issues adequately. This approach differs from existing data analytic pipelines that explore one sensing channel at a time. Integrating data types could fill the gap in digital health initiatives that include ANS data and thus address one of the major issues in this area.

\subsection{Addressing Reproducibility}
Scientific endeavors are experiencing both a crisis of method reproducibility (the ability to achieve congruent results from a given dataset and analysis) and results reproducibility (the ability to recreate results independently~\cite{goodman:Whatdoesresearch-2016, mishra:stress-reproducibility}). Large-scale projects indicate various degrees of non-replicability in multiple scientific fields~\cite{camerer:Evaluatingreplicabilitylaboratory-2016,sorkin:ChallengeReproducibilityAccuracy-2016,freedman:Impactpreclinicalirreproducibility-2015, opensciencecollaboration:Estimatingreproducibilitypsychological-2015, hutson:Artificialintelligencefaces-2018}. In addition, reported results are frequently incorrect or misstated~\cite{bakker:Misreportingstatistical-2011}. Some commonly reported reasons include a lack of interoperability between software tools and data sets and the limited record-keeping maintained for complicated data sets, which impact coordinated analyses, reporting, and archiving. For example, crowdsourced analysis projects where multiple expert teams analyze the same data corpus reveal an enormous amount of analytical flexibility present in complex analyses of identical questions (e.g. 29 teams analyzed an identical dataset, with odds ratios for effects ranging from $0.89$ to $2.93$, $M=1.31$~\cite{silberzahn:ManyAnalystsOne-2018}). Reproducibility of analyses require identical statistical analyses. 

The reproducibility crisis is pronounced in the context of ANS data collection and analysis when using consumer sensors in real-world settings. The field lacks clear and standardized guidelines for analyzing autonomic data, leading to many disparate methods and a bottleneck in translating promising proofs-of-principle to widespread use. Researchers repeatedly build new algorithms and methods from scratch on new datasets without any meaningful comparisons with existing approaches or different datasets. This lack of benchmarking leads to a cycle of ``reinventing the whee'' and ``demonstrating feasibility.''

\section{Envisioning an Open-Source, Community-driven Peripheral Psychophysiological Data Processing Framework}

We envision an open-source framework that enables community users to create, execute, and share computational models and data analysis pipelines that address standardization, interpretation, and reproducibility challenges often encountered when analyzing ANS data. While several free and open-source software platforms are available for peripheral physiological data analysis~\cite{kleckner:SimpleTransparentFlexible-2018,nabian:Biosignalspecificprocessingtool-2017,bach:Modelbasedanalysisskin-2013,kaufmann:ARTiiFACTtoolheart-2011,benedek:Decompositionskinconductance-2010, bartels:SinusCoradvancedtool-2017, perakakis:KARDIAMatlabsoftware-2010, pichot:HRVanalysisFreeSoftware-2016, mali:MatlabbasedtoolECG-2014, tarvainen:KubiosHRVHeart-2014,ramshur:DesignEvaluationApplication-2010,vollmer:Robustsimplereliable-2015,wagner:PhysiologicalSignalsEmotions-2005,silva:OpensourceToolboxAnalysing-2014, foll:FLIRTfeaturegeneration-2021}, including some well-known, feature-rich software packages (e.g., WFDB~\cite{silva:OpensourceToolboxAnalysing-2014} and openANSLAB~\cite{blechert:ANSLABIntegratedmultichannel-2016}), we identified 10 common problem areas that impede broader acceptability and usability: (i) focus on only one or a few biosignals, each requiring its own analysis pipeline and signal-specific expertise; (ii) often download-on-request, or ‘freemium’ (i.e., requiring payment for some analysis and input/output functions); (iii) designed for smaller laboratory datasets, not ambulatory datasets, which are typically much larger and require a significant investment of personnel time in error detection and correction; (iv) no integration across biosignals; (v) no explicit support for open scientific principles or platforms; (vi) unsupported, providing little or no accompanying documentation; (vii) inability to analyze the context in which data were collected; (viii) static, and not designed to incorporate code from other contributors; (ix) no ability to archive analysis pathways; and (x) command-line based, thus challenging for non-programmers. Proprietary software packages accompanying proprietary hardware and more general biosensor synchronization software have similar problems, as well as higher costs and closed code.

We developed a Survey of User Needs (SUN) to assess whether researchers and engineers from various scientific fields and disciplines who regularly process and analyze peripheral physiological and contextual data also experience the aforementioned analytical barriers. We circulated the survey to a large sample of researchers and engineers ($n=421$; $31\%$ Researchers, $69\%$ Engineers) from the Society of Psychophysiological Research, the IEEE International Machine Learning for Signal Processing Workshop, and snowball sampling using personal contacts and social media. Consistent with our review of existing software, over $70\%$ of the researchers we surveyed confirmed facing difficulties syncing data from different sources, identifying errors in data, and combining data from different types of devices. They also reported that the various software tools were hard to learn and lacked clear instructions. 

Our review of existing software and our survey results demonstrate that researchers and engineers across several disciplines working with multiple peripheral physiological signals would benefit from a multimodal data fusion platform with an open-source codebase. Indeed, SUN respondents were enthusiastic about a community-driven open-source framework that enables more transparency and reproducibility in the field. Based on this feedback, we outline the various core components we envision are needed.

\subsection{Community Driven}
We envision that the sustainability of the framework can be achieved by inviting scientists and researchers to contribute state-of-the-art methodologies and algorithms as plugins. Researchers can contribute individual tasks like artifact removal or complete end-to-end pipelines and machine-learning models for a particular outcome. Such a collaborative approach will be crucial in addressing reproducibility challenges by (a) allowing engineers and computer scientists to validate and refine their approaches and algorithms and (b) allowing behavioral scientists and clinicians access to cutting-edge tools and methods for their work. By integrating these community-contributed plugins, we envision a dynamic and ever-evolving platform, always up to date with the latest advancements in the field. Furthermore, all generated syntax, results, visualizations, and meta-data should be documented. These could be stored locally and on a networked storage system available via a public interface to promote transparency and reproducibility. We envision that every step (with version control) associated with data processing and analysis, which we call the data supply chain~\cite{goldsack:Verificationanalyticalvalidation-2020}, would be automatically saved as meta-data associated with a given dataset so other researchers can understand exactly how the data were collected, cleaned, and analyzed. 

\subsection{Data Quality Auditing and Preprocessing}
The framework should audit and assess the quality of peripheral physiological data collected in various settings by identifying and helping researchers address challenges like motion artifacts, environmental factors, and hardware limitations. To this end, we propose that the framework implement multiple semi-automated modules for data cleaning, preprocessing, and artifact removal, employing statistical and state-of-the-art machine-learning techniques as plugins~\cite{taylor:eda-explorer2015, gashi:DetectionArtifactsAmbulatory-2020, kelsey:Artifactdetectionelectrodermal-2017}. These software elements should cater to different data types and help researchers efficiently prepare and process their data for subsequent analysis.

\subsection{Signal Segmentation and Alignment}
A core aspect of ANS signal analysis is effective segmentation. Accordingly, the proposed framework should include a module that can determine appropriate time windows for signal segmentation based on the type of biosignal and research question. This module would help researchers and users select the optimal window length for their specific research needs, like short windows for time-domain features or longer ones for frequency-domain features. Aligning multimodal signals can be a challenge, particularly for uneven sampling rates, and tools can be included to support improved signal alignment~\cite{jiang:EventDTWImprovedDynamic-2020}.

\subsection{Contextual Information Integration}
Appropriately interpreting physiological data requires understanding the recording context, which includes characteristics of the environment outside the person (social, geolocation, ambient temperature, etc.) and those of the person (e.g., affect, physical activity, posture) that impact it. For instance, prior work demonstrates improved stress detection capabilities when physiological signals include contextual features~\cite{mishra:context}. While several libraries allow researchers to do contextual processing (e.g., BeWie, RAPIDS), none integrate contextual information to drive physiological data processing, i.e., provide users with plots of physiological data with visual overlays that describe recording context (location change, physical activity, etc.) to help researchers determine whether contextual feature variables should be controlled for in subsequent analyses. 

\subsection{Data Fusion and Signal Alignment}
Researchers often use multiple peripheral physiological sensors and signals to study a particular outcome. Thus, data fusion and signal alignment are crucial. This step involves aligning data from different sensors or modalities, which might vary in sampling frequencies and timestamps. We envision community-contributed plugins capable of handling the intricacies of multimodal physiological data by harmonizing signals across different timescales and accommodating both short-term events and long-term trends.

\subsection{Programming Language and GUI}
In the SUN survey, almost all behavioral scientists noted the need for a graphical user interface (GUI) to interact with open-source physiological processing tools. Thus, to cater to diverse user backgrounds and varying programming skills, the framework should be accessible both through a user-friendly GUI and a command-line interface (CLI). If common open programming languages (Python, R, Bash, etc.) are used, it would be convenient for researchers also to contribute their plugins and machine-learning models and use the framework to visualize their data and the outcomes of the various modules and plugins. 

\subsection{Science Gateways and Open Science Integration}
The envisioned framework should align with Open Science Framework standards, ensuring compatibility with open science practices. The integration could include leveraging the Digital Health Data Repository, where researchers can easily share open-sourced, de-identified datasets while following relevant data management and reporting standards and considering privacy and ethical constraints.

\section{DBDP Autonomic}
To effectuate our vision, we suggest expanding the Digital Biomarker Discovery Project (DBDP~\cite{bent:Digitalbiomarkerdiscovery-2020}) to include dedicated processing of ANS signals -- which we call DBDP Autonomic (\figref{dbdp-autonomic}). DBDP is designed to serve as a hub for collaborative and open research in the field of digital health. Its current code repository includes computational building blocks for the most common measures of ambulatory physiological data collected through wearable devices, including photoplethysmography (PPG), electrocardiography (ECG), and electrodermal activity (EDA). The repository comprises four modules: (1) exploratory data analysis, (2) data preprocessing, (3) feature engineering, and (4) machine learning model development. Together, they provide users with methods needed to complete each component of an end-to-end data processing pipeline, including data cleaning and preprocessing tasks, analysis, and predictive model development. 

In addition to these general method modules, DBDP hosts an archive of code repositories and a list of open-source digital health data from internal and external collaborators. Contributors can upload their code to the DBDP archive or actively collaborate to update the methods repositories. DBDP is also developing a code-free GUI-based platform (DBDP Discovery) that enables users with little or no coding expertise to interact with the functionalities of DBDP modules either with datasets from the DHDR or their own data appropriately formatted (CSV, excel, etc.)

\addfigure{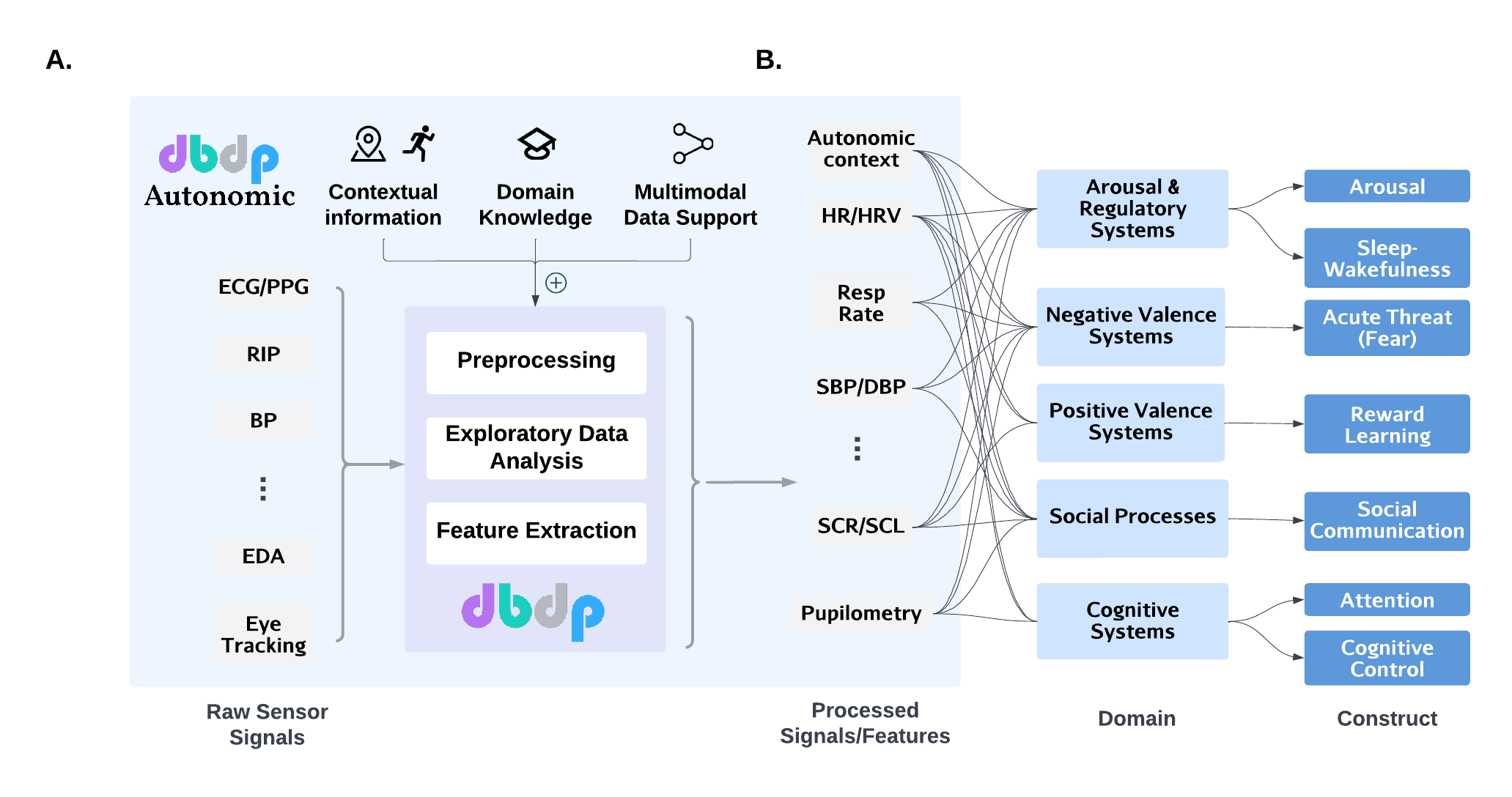}{1}{DBDP Autonomic extends the functionalities of DBDP for biobehavioral research. \textbf{A}. As signals from various modalities enter the analysis pipeline, DBDP Autonomic provides additional features on top of the existing modules in DBDP. These features extract and add contextual information, provide domain knowledge (for parameter tuning), and support multimodal data fusion. \textbf{B}. Researchers can then integrate the processed features such as HR, RR, and BP to understand the autonomic constructs in the context of major domains of basic human neurobehavioral functioning. \textit{ECG: electrocardiogram, PPG: photoplethysmography, RIP: respiratory inductance plethysmography, BP: blood pressure, EDA: exploratory data analysis, HR: heart rate, HRV: heart rate variability, SBP: systolic BP, DBP: diastolic BP, SCR:  skin conductance response, SCL: skin conductance level}. }

\section{Call to Community Action}
DBDP has established itself as an open-source hub for digital health, offering educational resources and computational tools for developing foundational features that collectively contribute to modeling complex biobehavioral outcomes. DBDP Autonomic, as an extension of this groundwork, could address the challenges associated with ANS signal analysis and enhance the standardization, interpretation, and reproducibility of existing and future research.

To advance biobehavioral research through DBDP Autonomic, we call upon the collective expertise of the digital health, behavioral, and psychophysiological research communities. Engagement and active participation from the community will be vital to ensuring the long-term viability and success of DBDP Autonomic. Our envisioned member engagement within DBDP Autonomic could follow the Center for Scientific Collaboration and Community Engagement (CSCCE) Community Participation Model (\figref{cscce-dbdp}), wherein multiple modes of interaction can coexist, with some members navigating through several nodes. Within this model, members typically initiate from the CONVEY/CONSUME mode, engaging with educational resources (e.g., tutorials and blog posts) to acquire biobehavioral and digital health knowledge. They may also access curated datasets and algorithms for their research. Transitioning to the CONTRIBUTE mode, research groups can add their go-to data cleaning algorithms, feature selection methods, and machine learning models as modules for other community members to use, benchmark, receive feedback, and cross-validate with other community-supplied methods. In the COLLABORATE mode, members of DBDP Autonomic could synergize and undertake joint research initiatives among the diverse community members who can apply, adapt, evaluate, and extend currently existing methods. Members could also co-author white papers and peer-reviewed publications to establish and enact standards in biobehavioral research. Lastly, members in the CO-CREATE mode could organize and lead workshops and working groups, driving the collective mission of DBDP Autonomic forward.

\addfigure{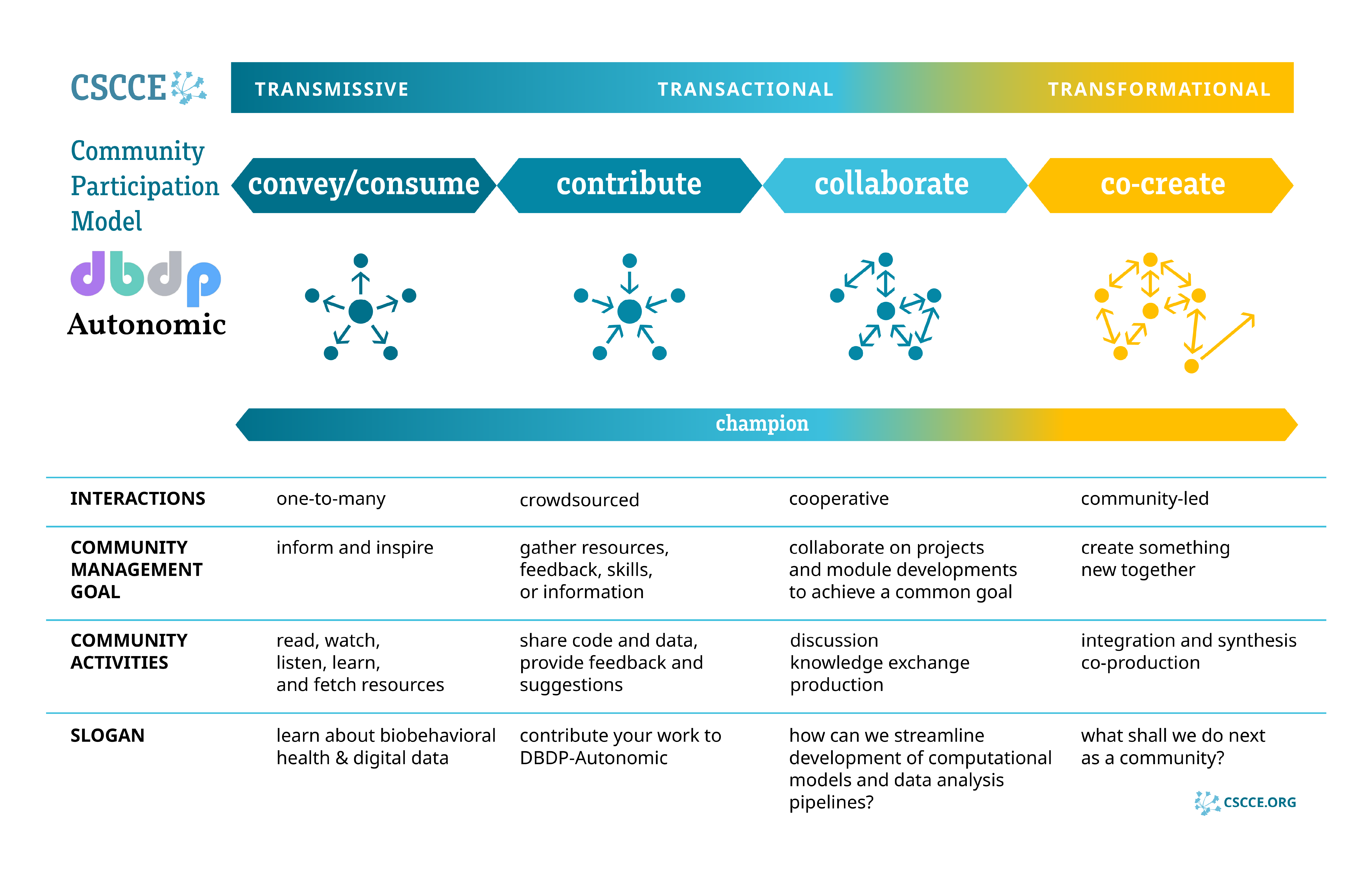}{1.0}{Community participation model of DBDP-Autonomic. DBDP-Autonomic offers diverse avenues through which users and collaborators can engage, from transmissive actions like `convey/consume' to transformational activities like `co-create.' Each stage caters to participants based on their unique goals, skills, and needs, allowing for multifaceted and inclusive community involvement, thus ensuring a richer and more holistic collaboration and contribution. This figure was adapted, with permission, from the Center for Scientific Collaboration and Community Engagement~\cite{woodley:CSCCECommunityParticipation-2020}.}

\section{Conclusion}
A collaborative effort of the DBDP Autonomic community could enable more robust, transparent, and reproducible research in biobehavioral health that involves ANS data. By emphasizing collaboration, transparency, and rigor, this resource could improve our understanding of complex biobehavioral health issues, provide personalized health insights, and accelerate the development of innovative interventions. As we move into the age of digital health, such a framework becomes essential for unlocking the full potential of mobile devices to benefit individual and community health.

%% file: ack.tex
\ifanonymized
 \relax
\else
 \section*{Acknowledgements}
 This research is partially supported by
the National Institutes of Health, under awards P30DA029926, 5RO1LM014191, R01DK133531; the National Science Foundation, under award 2339669; and the American Heart Association, under award 23POST1025599.   %
The authors would also like to thank the following colleagues for their insightful thoughts and discussions around this paper: Murat Akcakaya, Judith Andersen, Dana Brooks, Deniz Erdogmus, James Heathers, David Kaeli, Ian Kleckner, Sarah Ostadabbas, Karen Quigley, and Jolie Wormwood.
 The views and conclusions contained herein are those of the authors and should not be interpreted as necessarily representing the official policies, either expressed or implied, of the sponsors.
\fi %